\documentclass[doublecol,graphicx,times]{epl2} 
\usepackage{graphicx}
\usepackage{amsmath}
\usepackage{amsfonts}
\usepackage{amssymb}
\usepackage{amsbsy}

\newcommand{\sci}{Science}

\newcommand{\apj}{Ast\-ro\-phys\ J}

\newcommand{\grl}{Geo\-phys\ Res\ Lett}

\newcommand{\jgr}{J\ Geo\-phys Res}
\newcommand{\nat}{Nature}
\newcommand{\prl}{Phys\ Rev\ Lett}

\newcommand{\pop}{Phys\ Plasmas}

\def\|{{\sss\parallel}}

\newcommand{\w}[1]{\mbox{\boldmath{$#1$}}}

\def\sss{\scriptscriptstyle}
\title{Resistive scales for collisionless reconnection in space plasma}
\shorttitle{Collisionless reconnection} 

\author{R. A. Treumann\inst{1,2} \thanks{Visiting the International Space Science Institute, Bern, Switzerland}}
\shortauthor{R. A. Treumann}

\institute{ 
  \inst{1} Department of Geophysics and Environmental Sciences, Munich University, Munich, Germany\\                   
  \inst{2} Department of Physics and Astronomy, Dartmouth College, Hanover, NH 03755\\
  }
 

\abstract{We derive restrictions on collisional reconnection under the conditions in near-Earth space. The arguments are based on the precise definition and estimates of the resistive scale $L_\eta$ at the ion inertial site in collisionless reconnection, i.e. in the so-called `ion diffusion region' which is the region of Hall-current flow. 
}

\begin{document}

\maketitle
\section{Introduction}
Collisional magnetic reconnection is fast only in insulators. In vacuum, for instance, magnetic fields can reorder arbitrarily. Reconnection in a plasma of finite conductivity is a slow process, in contrast. In ideally conducting media, reconnection should be inhibited according to the frozen-in condition ${\w E}=-{\w v}\times{\w B}$ which is imposed on the medium by Lorentz-invariance and acts on the particles of charge $q$, velocity ${\w v}$ and kinetic energy $W$ in magnetic ${\w B}$ and electric ${\w E}$ fields. The frozen-in condition implies conservation of the magnetic moments $\mu=W_\perp/|{\w B}|$ of the charged particles tying them to the magnetic field. Undisputable observations {\it in situ} the collisionless (and therefore highly if not ideally conducting) space plasma have, on the other hand, proven for long time that reconnection is going on not only here as well but in addition at a very fast rate, indicating that reconnection is a kinetic process. Such observations refer to the magnetopause \cite{paschmann1979}, Earth's magnetotail \cite{fujimoto1997,nagai1998,oieroset2001} and most recently the solar wind \cite{phan2007}. Being fast means that reconnection proceeds on the ion or electron plasma spatial and time scales. Such observations have lead to attribute reconnection to various processes believed of having the potential to break the frozen-in condition by referring to the various different terms in the generalized Ohm's law \cite{krall1973,baumjohann1996,treumann2002} in plasma: generation of anomalous resistivity \cite{yamada2002,treumann2002,elkina2006}, extra-diagonal electron pressure terms \cite{hesse1995}, large-amplitude whistlers \cite{deng2001,rogers2001,pritchett2004}, the Hall-effect \cite{sonnerup1979,hoshino1996,fujimoto1997,asano2003,drake2008}, chaotic motions of ions and electrons \cite{buchner1986} in the current sheet separating oppositely directed magnetic fields, hard driving \cite{ramos2002}, magnetic-field aligned electric fields, ponderomotive forces \cite{treumann2002}, plasma composition etc. None of these processes has so far provided a satisfactory explanation of the occurrence of fast reconnection in ideally conducting plasma. Numerical simulations in two and three dimensions have, on the other hand, shown that reconnection indeed occurs under many different initial conditions. In particular, it also occurs in the complete absence of the Hall effect \cite{jaroschek2004} in pair plasma. However, in all these simulations reconnection is ignited externally in some particular way and is thus imposed on the plasma. For correctly chosen settings it then evolves and keeps itself going, allowing for the study of many secondary processes like magnetic topologies, generation of electric fields, current breakdown, wave excitation, anomalous diffusion \cite{elkina2006}, plasmoids, electron hole production \cite{drake2003,cattell2005}, plasma jetting, and particle acceleration \cite{drake2005,drake2006,jaroschek2009}. In this way numerical simulations in fluid, hybrid, kinetic Vlasov and full-particle codes as well as in two and three spatial dimensions have contributed substantially to understanding the properties of ongoing reconnection. They have not explained, however, the fundamental problem of why reconnection occurs under collisionless conditions. Fast reconnection in collisionless plasma indeed requires breaking the frozen-in condition. Here, on discussing the role of the so-called resistive scale, we investigate the conditions under which collisional reconnection in a warm dilute space plasma can be expected to occur, i.e. the conditions when it alone is capable of breaking the frozen-in condition.

\section{The resistive scale} On the scales of reconnection space plamas are (almost or nearly) collisionless. These scales are of the order of the ion inertial length $\lambda_i=c/\omega_{pi}$ or ion gyroradius $\rho_{ci}=v_i/\omega_{ci}$, where $c=(\mu_0\epsilon_0)^{-\frac{1}{2}}$ is the velocity of light, and $\omega_{pi}=(e^2N/\epsilon_0m_i)^\frac{1}{2}, \omega_{ci}=eB/m_i$ are the respective ion plasma and cyclotron frequencies in a plasma of density $N$ and magnetic field ${\w B}$ (with $e$ elementary charge and $m_i$ ion mass). Any resistivity must necessarily be anomalous, caused by wave particle interactions that are intrinsic to the plasma. This raises the question of how large this resistivity must be in order to be able to set the frozen-in concept out of action. To answer this question we refer to the magnetic induction equation
\begin{equation}\label{eq1}
\frac{\partial{\w B}}{\partial t}=\nabla\times{\w V}\times{\w B} +\frac{1}{\mu_0\sigma_a}\nabla^2{\w B} + \cdots
\end{equation}
where ${\w V}$ is the bulk flow velocity of the plasma, and $\sigma_a=e^2N/m_e\nu_{a}=\epsilon_0\omega_{pe}^2/\nu_a$ is the (anomalous) conductivity expressed through the square of the electron plasma frequency $\omega_{pe}$ and the anomalous collision frequency $\nu_a=v_e/\lambda_{a}$, which is a function of the electron thermal velocity $v_e$ and the anomalous mean free path $\lambda_a=(NS_a)^{-1}$, with $S_a$ the anomalous collisional cross section (a functional, e.g.,  of the scattering wave power). The dots $\cdots$ in Eq. (\ref{eq1}) stand for the various remaining terms in the generalised Ohm's law 
\begin{equation}\label{eq2}
\w{E + V\times B}= \eta_a{\w J} +\frac{1}{eN}\left(\w{J\times B}- \nabla\cdot{\textsf P}_e \right) + {\rm MCV + PM}
\end{equation}
Here $\eta_a=\sigma_a^{-1}$, the second term is the sum of the Hall-term $\w{J\times B}$ and the divergence of the electron thermal pressure tensor $\textsf{P}_e$. The last terms account symbolically for the mixed current-velocity (MCV)  \cite{krall1973,baumjohann1996} and ponderomotive (PM) terms \cite{treumann2002}. For the purposes of this Letter which intends to clarify the role of anomalous resistance all these terms are not considered here.

For comparing the relative importance of the two terms on the right of Eq. (\ref{eq1}) we introduce the convective and resistive scales $L_c$ and $L_\eta$, respectively. For the resistive term to dominate and destroy the frozen-in state of the plasma one requires that dimensionally
\begin{equation}
\mu_0\sigma_aL_\eta^2<L_c/V
\end{equation}
Since it can be shown that $\eta_a=\mu_0\nu_a\lambda_e^2$, with $\lambda_e=c/\omega_{pe}$ the electron skin depth,  this expression becomes 
\begin{equation}
L_\eta^2<\nu_a\lambda_e^2L_c/V=\nu_a\lambda_e^2\tau_c
\end{equation}
where $\tau_c$ is the time the flow needs to cross the reconnection site over the distance of $L_c$. Using the definition of the anomalous collision frequency this inequality can be brought into the following form
\begin{equation}
L_\eta^2/\lambda_e^2<v_e\tau_c/\lambda_{a} =\omega_{pe}\tau_c (\lambda_D/\lambda_a)
\end{equation}
The quantity $\lambda_D=v_e/\omega_{pe}$ is the Debye length.

The last expression can be brought into an applicable form if we refer to the observation that reconnection sets on under conditions when the width of the separating current layer becomes less than the ion inertial length scale $\lambda_i$, in which case the ions become non-magnetised, while the magnetic field remains to be tied to the electrons such that it is the electron flow across the ion inertial region which transports the magnetic field into the centre of the reconnection site to undergo reconnection. Still, during this transport, the electrons are frozen to the magnetic field such that the problem is shifted to a process that enables the electrons to leave the field in order enabling it to reconnect. For anomalous resistance to provide such a mechanism one thus needs to express $\tau_c$ in terms of $\lambda_i$ as 
\begin{equation}
\tau_c=L_c/V\sim \lambda_i/V= (c/V_A)(\omega_{pi}M_A)^{-1}
\end{equation}
where we introduced the Alfv\'enic Mach number $M_A=V/V_A<1$ expressing the velocity through the Alfv\'en velocity $V_A=B/\sqrt{\mu_0m_iN}$, and $c/V_A=\omega_{pi}/\omega_{ci}$. \{Usually, in reconnection, $M_A\sim {\rm O(10^{-2})}\ll 1$, except for some claims that reconnection might also go on in the shock front of high Mach-number (presumably relativistic) shocks \cite{gedalin2009} where, if true, it could substantially shorten the resistive scale.\} We can then write
\begin{equation}
L_\eta^2/\lambda_e^2\lesssim (m_i/m_e)(\omega_{pe}/\omega_{ce})(\lambda_D/\lambda_a)M_A^{-1}
\end{equation}
Finally, expressing $\lambda_D\omega_{pe}/\omega_{ce}=\rho_e$ through the electron gyroradius $\rho_e=v_e/\omega_{ce}$, we obtain for the resistive scale the following expression 
\begin{equation}\label{eq8}
\frac{L_\eta}{\lambda_e}\lesssim\left[\frac{1}{M_A}\left(\frac{m_i}{m_e}\right)^{\!\!\frac{1}{2}}\!\frac{\rho_i}{\lambda_a}\right]^\frac{1}{2}
\end{equation}
that is applicable to current sheets of widths $\Delta\lesssim \lambda_i$, smaller than the ion inertial length scale.

It is important to note that the resistive scale is not the same as the mean free path of the particles. The final expression Eq. (\ref{eq8}) gives the resistive scale in terms of the anomalous mean free path. The resistive scale refers to the length over that the magnetic field energy can resistively be dissipated converting magnetic energy into heating the plasma. Therefore, the resistive scale has to be defined in terms of the process that provides the dissipation. 

General arguments are based on the notion that such scales must be short in order to match the internal correlation lengths. In reconnection it is reasonable to assume that these lengths are related to the scales on which the particles become demagnetised allowing the magnetic field to slip away from the particles. Since in collisionless reconnection the ions in the Hall region are already non-magnetic and the transport of the magnetic field is provided by the electrons, the microscopic correlation scale for reconnection is quite naturally the electron inertial length $\lambda_e$. Once the resistive scale becomes the order of the electron skin depth $L_\eta\sim {\rm O}(\lambda_e)$, one expects that reconnection becomes anomalously resistive even though the plasma is collisionless. In the following we investigate the consequences of this assumption.

\section{Resistive reconnection in collisionless plasma - a myth?} It is frequently assumed that some anomalous resistance can be produced in collisionlesss plasma. We are now going to check whether anomalous resistivity can play a dominant role in collisionless reconnection. In the spirit of the above discussion we assume that some anomalous resistance has indeed be generated locally in the reconnection site. For reconnection to work one then in the inequality Eq. (\ref{eq8}) requires that $L_\eta\sim\lambda_e$, in which case the left hand side becomes unity and Eq. (\ref{eq8}) turns into an expression from which the anomalous mean free path length $\lambda_a$ can be determined as
\begin{equation}\label{eq9}
\lambda_a=\left(\frac{m_i}{m_e}\right)^{\!\!\frac{1}{2}}\frac{\rho_i}{M_A}
\end{equation}
Since inflow speeds $V$ in reconnection are small, usually much less than the Alfv\'en velocity, the Mach number $M_A$ is small suggesting that the anomalous mean free path can be quite large, indeed. 

For a more quantitative conclusion we have to compare this anomalous mean free path with the Coulomb-Spitzer-Braginskii (collisional) mean free path $\lambda_{\rm CSB}$ of the plasma. Instead it is more convenient to use the cross sections. The CSB-cross section can be expressed in terms of the Debye length as $S_{\rm CSB} \simeq \Lambda /16\pi N^2\lambda_D^4$ with $\Lambda$ the Coulomb logarithm (which corrects for small angle scattering). Note that  small cross sections imply weak scattering and thus  inefficient scattering processes!

The ratio of the two cross sections in question, on using $\lambda_D/\rho_i=(V_A/c)(m_i/m_e)^\frac{1}{2}$, is then
\begin{equation}
R_a\equiv\frac{S_a}{S_{\rm CSB}}\simeq \frac{16\pi}{\Lambda}\beta_rN\lambda_D^3
\end{equation}
where $\beta_r=V/c$ is the relativistic factor of the inflow, and $(N\lambda_D^3)^{-1}=W/NT$ is the thermal plasma wave fluctuation intensity $W$ which causes the Coulomb scattering, normalised to the thermal energy density $NT$ (with temperature $T$ in energy units). The latter is quite small in the dilute space plasma. Its inverse appearance in the last expression indicates that the anomalous scattering cross section must become very large in order to   let anomalous resistivity play a substantial role in reconnection in an otherwise collisionless plasma.

For an estimate we refer to the conditions in the tail plasma sheet of the Earth's magnetosphere both during quiet or substorm conditions. Inflow velocities are of the order of $V\sim$ a few 10 km/s yielding $\beta_r\sim 10^{-4}$ (or Mach numbers $M_A\sim 10^{-2}$). Plasma sheet densities are between
$10^5\lesssim N\lesssim 10^6$ m$^{-3}$. Electron temperatures are in the range $30\lesssim T_e\lesssim 100$ eV. These values yield cross section ratios 
\begin{equation}
10^8\lesssim R_a \lesssim 10^{11}
\end{equation}
with the smaller value holding under quiet magnetospheric conditions (small $V$ and $\lambda_D$). For the anomalous cross section to exceed the CSB cross section by so many orders of magnitude a rather strong instability is required. The CSB-collision frequency under these conditions is of the order of $\nu_{\rm CSB}\sim 10^{-9}$ s$^{-1}$. Since the collision frequency scales with the cross section, the above scaling range requires anomalous collision rates of the order of $\nu_a\sim (1-10)$ s$^{-1}$ which is of the order of  the lower hybrid frequency in a 10 nT magnetic field, presumably the largest possible value anomalous collision frequencies can reach under conditions of magnetised electrons and non-magnetised ions. Moreover, the anomalous mean free path from Eq. (\ref{eq9}) becomes $\lambda_a\sim 2.4\times10^5$ km, much smaller than $\lambda_{\rm CSB}$ but still much larger than $\lambda_i$. Thus, even though not impossible, it is highly improbable that reconnection in the geomagnetic tail current sheet will be caused by the generation of anomalous resistance. In fact, the required high wave fluctuation level of $W_a\sim 10^{10} W$ needed to reproduce the above value of $R_a$ has never been reported in observations of reconnection. In contrast, any observed wave levels in presumable crossings of the very reconnection site \cite{treumann1990,bale2002} have been found to be astonishingly low both in the magnetospheric tail plasma sheet as at the dayside of the magnetopause. Thus anomalously collisional reconnection in the dilute near-Earth space plasma is probably a myth.  

\begin{figure}[t]
\centerline{{\includegraphics[width=8.6cm]{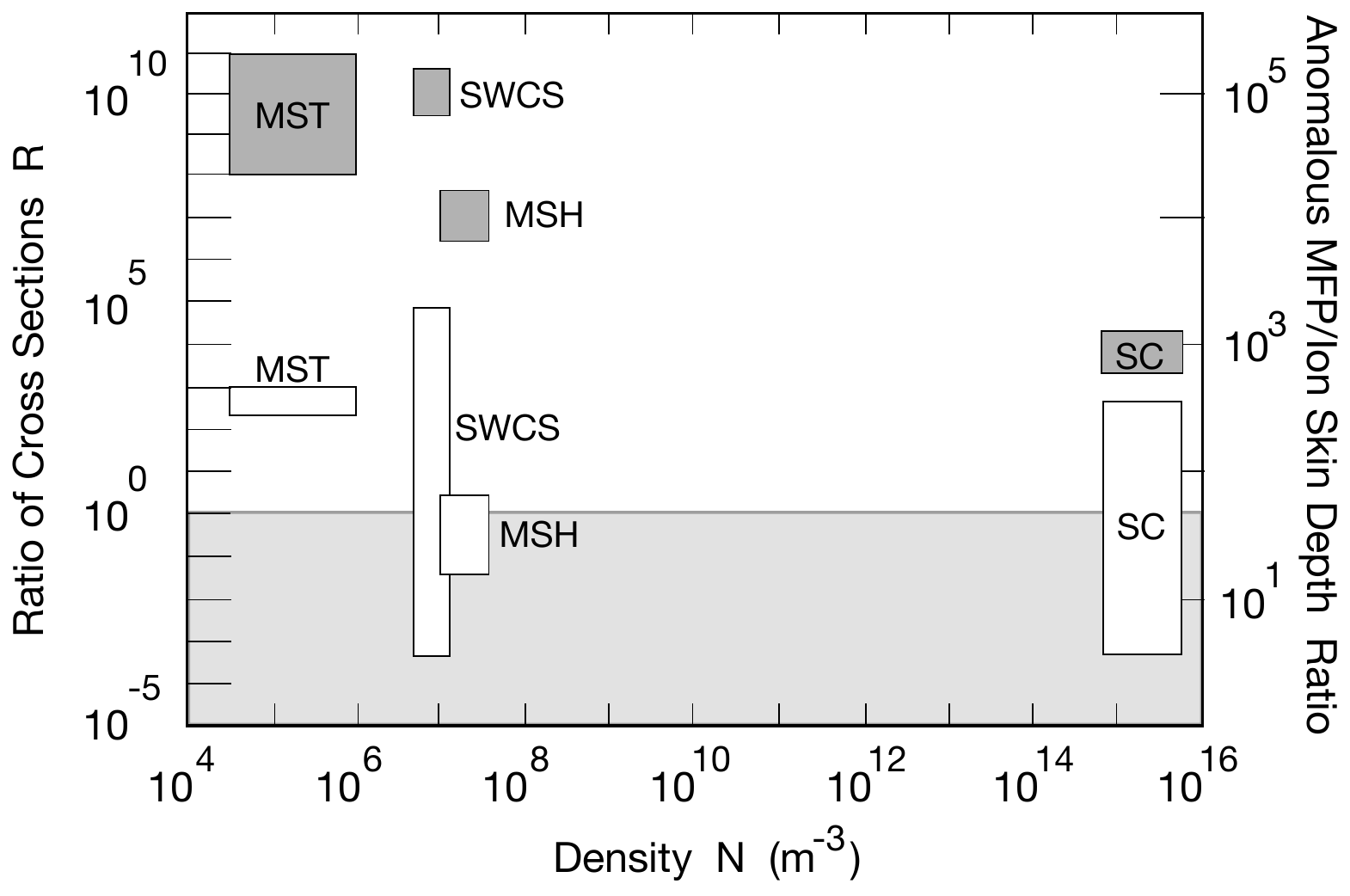}}}
\caption{The ratios $R=S_a/S_{\rm CSB}$ (filled rectangles) and $\lambda_a/\lambda_i$ (white rectangles) in the Solar Corona (SC), Earth's Magnetosheath (MSH), Earth's Magnetospheric Tail (MST) and Solar Wind Current Sheet (SWCS) in dependence on the plasma density in these regions. The shaded area below the line $R=1$ indicates the parameter region for $R$ where the Coulomb-Spitzer-Braginskii cross-section is larger than the anomalous cross section in the ion inertial region. There is some variation in the local parameters of each region which is approximately accounted for in the size of the rectangular regions. In particular, the local Mach numbers entering $\lambda_a$ are relatively uncertain. In all regions under consideration $R\gg1$. The ratios $\lambda_a/\lambda_i$ for SC and SWCS are quite uncertain but their range still exceeds 1.}\label{recmyth-fig}
\end{figure}

\section{Discussion} Since reconnection has been unambiguously identified in space from its many secondary signatures and since, as we have concluded here from a purely theoretical investigation of scales, it would be difficult to realise the required small resistive scales based on anomalous resistance produced by wave-particle interaction at the reconnection side, reconnection in collisionless space plasmas is caused by other effects than anomalous collisions. The generalised Ohm's law Eq. (\ref{eq2}) still allows for a number of different actors, some of which may be capable of breaking the frozen in condition on the left hand side of Eq. (\ref{eq2}). Since the Hall force itself does not dissipate energy, the Hall term, inserted into the induction law Eq. (\ref{eq1}), may still act through the density gradient term $(\w{JB-BJ})\cdot\nabla(eN^{-1})$ to break the frozen-in state. Similarly, the non-gradient part of the pressure tensor divergence provides another possibility, as do the two remaining terms. In particular the ponderomotive force term is of interest. It is non-zero when the turbulent wave field in the ion-inertial range consists of localised structures like electron holes which locally possess high wave intensity. As long as the ponderomotive force is derived from a gradient of wave power, as is the usual case in solitons like ion-acoustic solitons, Langmuir solitons and so on (cf., e.g., \cite{treumann1997}) it will, like the Hall term, contribute only through the inverse density gradient factor. However, when it possesses genuine non-gradient terms (as probably happens in Bernstein-Green-Kruskal modes), then breaking the frozen-in condition may be likely due to the ponderomotive force. We may hence conclude that anomalous resistance alone is very unlikely to provide sufficient dissipation in the plasma of near-Earth space for acting as the sole cause of reconnection. However, a weak resistivity on a resistive scale that is substantially larger than the electron scale may help, together with other effects that are contained in the generalised Ohm's law, to keep reconnection going even though anomalous resistivity by itself does not suffice for keeping it going.

Let us finally check the variation of the anomalous cross section in different regions of space. In the solar corona (SC) one has $N\sim 10^{15}$ m$^{-3}$ (at the $\sim$ 300 MHz radio emission level), $T_e\sim 10^2$ eV, $B\sim (10^{-4}-10^{-1})$ T (the larger value probably approximately valid in active magnetic regions during flares), yielding $\lambda_D\sim2\times 10^{-3}$ m, $\lambda_e\sim 0.1$ m, $\lambda_i\sim 4$ m, $N\lambda_D^3\sim 1.3\times 10^{7}$, and an Alfv\'en velocity between $70\lesssim V_A\lesssim 7\times10^4$ km/s. Typical flow velocities $V$ are in the range of several $10\lesssim V\lesssim$ several 100 km/s. The lower hybrid frequency is in the range $7\,{\rm kHz}<f_{lh}<1$ MHz, and the anomalous mean free path lengths range in between $15\,{\rm m}\lesssim \lambda_a\lesssim 1.5$ km. These values suggest a cross section ratio $3\times 10^{3}<R_a<2\times10^{4}$ for anomalous processes becoming important. It shows that in the corona the required anomalous cross section is also substantially larger than the classical cross section $S_{\rm CSB}\sim 4\times10^{-19}$ m$^2$, corresponding to a collisional mean free path of $\lambda_{\rm CSB}\sim 3$ km. The anomalous mean free path needed for reconnection to become resistive is $0.02/M_A< \lambda_a<15/M_A$ m. Since the inflow velocities are probably much less than the Alfv\'en speed, though little is known in this respect, $M_A$ can range from $M_A>10^{-3}$ to $M_A\lesssim 1$ which for $\lambda_a$ implies that $20\,{\rm m}<\lambda_a$. This is larger than $\lambda_i$ which in spite of the above conclusion suggests that in the solar corona the conditions on anomalous resistance are less restrictive than in the magnetosphere. 

In the solar wind current sheet (SWCS) the parameters are $N\sim 10^7$ m$^{-3}$, $T_e\sim 10^2$ eV, $B\sim (1-5)$ nT, $V_A\sim (10-20)$ km/s, $\lambda_e\sim 2$ km, $\lambda_i\sim 70$ km, $\lambda_D\sim 23$ m, yielding $N\lambda_D^3\sim 1.3\times10^{11}$. Flow velocities across the sheet are quite uncertain but can at most be of the order of the solar wind flow speed $V<10^3$ km/s. Hence the required relative anomalous cross section is in the range $R_a< 10^{10}$. The required anomalous mean free path becomes $\lambda_a\sim (2.4/M_A)\times10^3$ km. Its dependence on the Mach number forces one to distinguish between convected (floating) and standing solar wind current sheets. The latter experience a substantial fraction of the solar wind Mach number $M_A\sim 8$, wile the former see only slow inflows with Mach number $0.01<M_A<0.1$. Hence $\lambda_a\sim 300$ km for standing and $2\times10^4<\lambda_a<2\times10^5$ km for floating current sheets, all substantially less than the ion inertial length thus requiring strong anomalous effects the realisation of which is hard to believe. 

Turning to the magnetosheath (MS) where reconnection has been suggested to occur \cite{retino2007} and play a role in dissipation \cite{sundqvist2007} we have $N\sim 3\times10^7$ m$^{-3}$, $T_e\sim 30$ eV, $B\sim 30$ nT, $V_A\sim (30-100)$ km/s, $\lambda_e\sim 1$ km, $\lambda_i\sim$ 40 km, $\lambda_D\sim 7$ m. Flow velocities across the narrow current sheets are at most of the order of some $V\sim 10$ km/s. This yields a required ratio $R_a\sim 10^{7}$ and anomalous mean free path $\lambda_a\simeq 70/M_A$ km.  One might assume that a maximum $M_A\sim 1$ behind the shock in the magnetosheath corresponding to a maximum flow speed, but probably the relative inflow in any of the small scale current sheets that have been detected here is much slower yielding $M_A\sim$ 0.1-0.2 only and thus required anomalous mean free paths $\lambda_a\sim 10^3$ km. Since all these mean free path estimates in the magnetosheath exceed the current sheet widths $\Delta\sim \lambda_i$ required in collisionless reconnection is is unlikely that anomalous collisional transport  processes contribute to reconnection here. Processes of generating anomalous diffusion that do not meet these values can thus be put aside as being involved in maintaining anomalous collisional reconnection in the magnetosheath. Figure \ref{recmyth-fig} gives an impression of the locations of the ratios $R_a$ and $\lambda_a/\lambda_i$ in the four regions investigated in this Letter.

In summary, we have tried clarifying the notion of the resistive scale in application to reconnection in those space plasma regions that are accessible to measurements {\it in situ} in near-Earth's space, finding that, under the conditions that prevail in the ion inertial region, this scale can hardly be made small enough for anomalous resistivity to drive reconnection as the sole responsible process in fast magnetic merging.

{\small \acknowledgements
This research is part of a Visiting Scientist Programme at ISSI, Bern. Hospitality of the ISSI staff and directors is thankfully recognised.  

}

\end{document}